\documentclass[namedreferences]{SolarPhysics}
\usepackage[optionalrh,solaenum]{spr-sola-addons} % For Solar Physics
\usepackage{graphicx}        % For eps figures, newer & more powerfull
\usepackage{color}           % For color text: \color command
\usepackage{url}             % For breaking URLs easily trough lines
\newcommand{\beq}{\begin{equation}}
\newcommand{\eeq}{\end{equation}}
\newcommand{\bfig}{\begin{figure}}
\newcommand{\efig}{\end{figure}}
\newcommand{\ben}{\begin{enumerate}}
\newcommand{\een}{\end{enumerate}}

\begin{document}
\begin{article}
\begin{opening}

\title{Nonlinear force-free reconstruction
of the global solar magnetic field: methodology}

\author{I.~\surname{Contopoulos}\sep
        C.~\surname{Kalapotharakos}\sep
        M.~K.~\surname{Georgoulis}}
\runningauthor{Contopoulos, Kalapotharakos, Georgoulis}
\runningtitle{Global Solar Magnetic Field}

   \institute{Research Center for Astronomy and Applied
     Mathematics (RCAAM), Academy of Athens,
     4 Soranou Efessiou Str., Athens 11527, Greece,
     email: \url{icontop@academyofathens.gr}}
\begin{abstract}
We present a novel numerical method that allows the calculation of
nonlinear force-free magnetostatic solutions above a boundary
surface on which only the distribution of the normal magnetic
field component is given. The method relies on the theory of
force-free electrodynamics and applies directly to the
reconstruction of the solar coronal magnetic field for a given
distribution of the photospheric radial field component.
%It can also
%self-consistently account for the line-of-sight correction
%required for magnetic field measurements taken away from the
%center of the solar disk.
The method works as follows: we start with any initial
magnetostatic global field configuration (e.g. zero, dipole), and
along the boundary surface we create an evolving distribution of
tangential (horizontal) electric fields that, via Faraday's
equation, give rise to a respective normal field distribution
approaching asymptotically the target distribution. At the same
time, these electric fields are used as boundary condition to
numerically evolve the resulting electromagnetic field above the
boundary surface, modelled as a thin ideal plasma with
non-reflecting, perfectly absorbing outer boundaries. The
simulation relaxes to a nonlinear force-free configuration that
satisfies the given normal field distribution on the boundary.
This is different from existing methods relying on a fixed
boundary condition - the boundary evolves toward the a priori
given one, at the same time evolving the three-dimensional field
solution above it. Moreover, this is the first time a nonlinear
force-free solution is reached by using only the normal field
component on the boundary. This solution is not unique, but
depends on the initial magnetic field configuration and on the
evolutionary course along the boundary surface. To our knowledge,
this is the first time that the formalism of force-free
electrodynamics, used very successfully in other astrophysical
contexts, is applied to the global solar magnetic field.
\end{abstract}
\end{opening}
%-------------------------------------------------

\newpage
\section{Introduction}
     \label{S-Introduction}
Force-Free Electrodynamics (hereafter FFE) is a formal name for
time-dependent electromagnetism in an ideal plasma with negligible
inertia ($\rho=0$) and negligible gas pressure ($\beta=0$). The
formalism of FFE has been developed for various relativistic
astrophysical applications (pulsars, astrophysical jets, gamma-ray
bursts, etc.) where the plasma supports electric currents and
electric charges \cite{gruzinov99}. The equations of FFE are
Maxwell's equations with nonzero electric fields as follows:
\begin{equation}
\frac{\partial {\bf E}}{\partial t} = c \nabla\times {\bf B} -
4\pi {\bf J} \ , \label{M1}
\end{equation}
\begin{equation}
\frac{\partial {\bf B}}{\partial t} = -c \nabla\times {\bf E}\ ,
\label{M2}
\end{equation}
\begin{equation}
\nabla\cdot {\bf B}=0\ .\label{M3}
\end{equation}
These equations are coupled by the ideal MHD condition
\begin{equation}
{\bf E}\cdot {\bf B} = 0 \ , \label{idealMHD}
\end{equation}
implying that in an ideal plasma ${\bf E}$ and ${\bf B}$ are
everywhere perpendicular, and the force-free condition
\begin{equation}
\rho_e {\bf E}+\frac{1}{c}{\bf J}\times {\bf B} = 0\ .
\label{forcefree}
\end{equation}
Here, ${\bf J}$ is the electric current density, and $\rho_e\equiv
(4\pi)^{-1}\nabla\cdot {\bf E}$ is the electric charge density.
\inlinecite{gruzinov99} showed that it is possible to solve for
${\bf J}$ in the above set of equations and thus express the
electric current density as a function of the electric and
magnetic field, namely
\begin{equation}
{\bf J} = \frac{c}{4\pi}\nabla \cdot {\bf E}\ \frac{{\bf E}\times
{\bf B}}{B^2} + \frac{c}{4\pi}\frac{({\bf B}\cdot \nabla\times
{\bf B} - {\bf E}\cdot \nabla\times {\bf E})}{B^2}\ {\bf B}\ .
\label{J}
\end{equation}
One can then numerically integrate Maxwell's equations
(eqs~\ref{M1}, \ref{M2}) to obtain the time evolution of the
electric and magnetic fields. We developed a three-dimensional
(hereafter 3D) finite-difference time-domain (FDTD) cartesian FFE
code with non-reflecting, perfectly absorbing outer boundaries
applied successfully to the 3D structure of the pulsar
magnetosphere (\opencite{kc09}; see also \opencite{yee66};
%\opencite{th05};
\opencite{spitkovsky06}).

We then realized that the same formalism may be applied to follow
the temporal evolution of any open force-free ideal-MHD system
from an initial to a final magnetostatic configuration (i.e. with
zero initial and final electric fields) when the magnetic field
distribution at a boundary evolves for a finite number of
time-iteration steps. After sufficient time, when the magnetic
field at the boundary will have reached a given distribution, and
all electric fields will have relaxed to zero, the above
expressions degenerate into the following limited set of
steady-state FFE equations,
\begin{equation}
{\bf J}= \frac{c}{4\pi}\nabla\times {\bf B}\ ,\label{L1}
\end{equation}
\begin{equation}
\nabla\cdot {\bf B}=0\ ,\ \mbox{and}
\end{equation}
\begin{equation}
{\bf J}\times {\bf B}=0\ .\label{forcefree2}
\end{equation}
A solution of the system of eqs.~(\ref{L1})-(\ref{forcefree2})
satisfying a given normal-field distribution on the boundary can
be viewed as the final FFE equilibrium solution. Given the
ill-posed nature of this problem (only the normal field component
on the boundary is known), this solution is not unique and depends
on the initial magnetostatic configuration and on the course
toward equilibrium\footnote{The above statements are correct
mathematically. In practice, however, as shown in \S~3, numerical
dissipation sets in and the system evolves adiabatically through a
sequence of nonlinear force-free magnetostatic equilibria with
decreasing magnetic free energy towards a unique, minimum-energy
and hence current-free (potential) equilibrium.}.

The above methodology can be applied to the inner solar corona,
that is, between 1 and $2-2.5 r_{\odot}$ (solar radii). In this
environment, motions in the line-tying photosphere largely but
nonlinearly dictate the evolution in the magnetic field lines
above, along which plasma is assumed `frozen in' and forced to
follow their path ($\beta\ll 1$). At the edge of the inner corona,
known as the `source surface' \cite{schatten69} the plasma again
approaches equipartition with the magnetic field. Beyond the
source surface the plasma starts dominating again in the form of
the solar wind. Therefore, between $1 r_{\odot}$ (photosphere) and
$\sim 2.5 r_{\odot}$ (source surface) the magnetic field is
thought to approximately satisfy
eqs.~(\ref{L1})-(\ref{forcefree2}) along with the divergence-free
condition, eq.~(\ref{M3}), and the ideal MHD condition,
eq.~(\ref{idealMHD}). In discussing the global solar field, and
driven by the necessity to use the observed photospheric magnetic
field as the boundary condition, we ignore the fact that the
latter is, in fact, forced, rather than force-free
\cite{metcalf_etal95,georgoulis_labonte04}.

Routine measurements of the global photospheric field provide only
the line-of-sight magnetic field component $B_{LOS}(r_{\odot},
\theta, \phi)$ for given heliocentric spherical coordinates
$(\theta,\phi)$\footnote{The Vector SpectroMagnetograph (VSM;
\opencite{henney_etal09}) of the Synoptic Optical Long Term
Investigations of the Sun (SOLIS) facility and the Helioseismic
and Magnetic Imager (HMI) onboard the Solar Dynamics Observatory
(SDO) is and will be providing, respectively, the first full-disk
photospheric vector magnetograms.}. This is typically accomplished
by synoptic (Carrington) maps of $B_{LOS}$ where the heliographic
latitude and the Carrington longitude can be transformed into
$\theta$ and $\phi$, respectively. Synoptic maps take one solar
rotation ($\sim 27$~days at the equator) to complete one
Carrington rotation ($0^\circ$ - $360^\circ$ in Carrington
longitude) but are again assumed to capture a snapshot of the
photospheric magnetic field for practical reasons (see, however,
the attempt to evolve the photospheric field by means of a flux
dispersal model \cite{schrijver_derosa03} and use the modified
boundary for global field extrapolations).

Determining the distribution of coronal magnetic fields and
subsequent electric currents is, per the above, a formidable task.
Global solar field extrapolations are current-free (potential) in
their majority
\cite{altschuler_newkirk69,wang_sheeley92,luhmann_etal02,schrijver_derosa03}.
Some magnetohydrodynamic modelling efforts are also present in the
literature \cite{linker_etal99,riley_etal01,roussev_etal03} but
they require immense computing resources. To our knowledge, the
only technique providing a global solar nonlinear force-free field
is the optimization method of T. Wiegelmann and collaborators. A
magnetostatic version of this method was proposed by
\inlinecite{wiegelmann_07}. The main idea was to simultaneously
minimize the Lorentz force ($\nabla \times {\bf B}) \times {\bf
B}$ and the divergence ($\nabla \cdot {\bf B}$) of the field in
the extrapolation volume in spherical coordinates - a
generalization of the Cartesian implementation of
\inlinecite{wiegelmann_04}. Efforts to further refine the method
are discussed in \inlinecite{tadesse_etal09} while a further
generalization into a magnetohydrostatic nonlinear force-free
solution is described by \inlinecite{wiegelmann_etal07} and
\inlinecite{ruan_etal08}. These optimization efforts, however,
generally require the vector magnetic field on the photospheric
boundary. We hereby propose an alternative approach reaching a
nonlinear force-free magnetostatic solution by only using the
normal (radial) field component on the boundary, as follows:
\begin{enumerate}
\item Start with a simple static potential magnetic field
distribution such as ${\bf B}=0$ everywhere, or a dipolar field of
any direction, etc. \item Evolve the vertical photospheric
magnetic field by imposing a distribution of horizontal
photospheric electric fields using Faraday's equation
(eq.~\ref{M2}). This distribution is chosen such that the
resulting photospheric magnetic field asymptotically
($t\rightarrow \infty$) approaches a given photospheric
magnetogram (see next section). \item During the evolution of the
photospheric magnetic field, force-free electrodynamic waves are
injected from the photosphere into the corona. Assuming
non-reflecting, perfectly absorbing conditions at large distances,
these waves will be absorbed by the outer boundaries and will not
re-enter the computational domain. The moving electric charges
carried by these waves begin to establish a nonlinear network of
coronal electric currents. \item Gradually, as the photospheric
magnetic field distribution approaches the target, photospheric
electric fields will correspondingly decrease, and as a
consequence magnetospheric electrodynamic waves, electric fields,
and electric charges will also asymptotically fade. The nonlinear
network of coronal electric currents, however, will survive. \item
When electric fields vanish everywhere, we relax to a force-free
configuration that closely matches the given photospheric boundary
condition.
\end{enumerate}
At this initial stage we cannot claim that our solution is
representative of the actual global coronal field. In fact, and as
argued above, our solution depends on the initial condition and on
the course followed to reproduce the given boundary condition
(see, however, footnote~1). To achieve a physically meaningful
equilibrium solution we need to apply additional physical
arguments that will be the subject of a future publication. This
work simply lays out the details of the methodology and presents a
few characteristic test cases. Notice that our methodology {\em
does not} apply in studying the dynamical evolution of the solar
corona between successive photospheric configurations (the
characteristic speed of propagation of information in the solar
corona, the Alfv\`{e}n speed, is on the order of 1,000~km/sec,
whereas the characteristic speed of propagation of information in
our electrodynamic `corona', the speed of light, is $\sim 300$
times faster). It does apply, though, in determining the
equilibrium coronal configuration.

Our method has certain similarities with `stress-and-relax'
schemes proposed by e.g. \inlinecite{1996ApJ...473.1095R},
\inlinecite{2005A&A...433..335V}, in that we both act on the
boundary magnetic field distribution, and then evolve the
magnetospheric field towards stationarity. Our respective
technical implementations, though, are different (e.g. they work
with the vector potential, we work only with the magnetic field;
they `stress' the horizontal boundary magnetic field component, we
evolve the radial one; they rely on viscous damping in order to
settle to stationarity, we implement absorbing outer boundary
conditions that remove any non-stationarity from our computational
box).

In \S~2 we present our numerical implementation of a photospheric
electric field distribution which guarantees that the distribution
of the radial photospheric magnetic field component will
asymptotically approach that of a given solar magnetogram. In \S~3
we follow the electrodynamic response of the corona under the
action of the above photospheric electric field distribution, and
present some representative solutions for various initial
conditions. We also present the various quantities used to monitor
the course toward equilibrium.
%Among them are monitors of the
%force-free and the divergence-free conditions.
Our conclusions are summarized and discussed in \S~4.

\section{The photospheric magnetic field}
\label{S-phot}
\begin{figure}[t]
\centerline{\includegraphics[width=.5\textwidth,clip=]{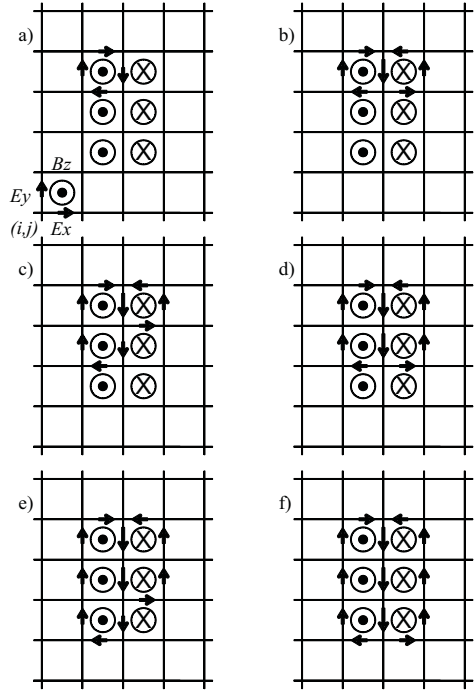}}
\caption{Implementation of electric field update algorithm
(eqs.~\ref{scanning}) at $t=0$ in a 2D staggered cartesian grid.
Electric fields are defined along grid edges. Magnetic fields are
defined in cell centers. Nonzero equal-magnitude target magnetic
field $B_T(i,j)$ only at positions shown with circles. Circles
with dots: $B_T$ outwards from the plane. Circles with crosses:
$B_T$ into the plane. The grid is scanned left to right, top to
bottom in frames a-f respectively.} \label{2Dalgorithm}
\end{figure}

Let us first assume a local 2D staggered cartesian grid $(x,y)$ of
size element $\delta$ in the photosphere, with normal magnetic
field $B_z$ defined at the center of each cell and tangential
electric fields $E_x, E_y$ defined on the corresponding cells'
edges, as shown in Fig.~\ref{2Dalgorithm}a. Each cell is
characterized by a vector position $(i,j)$. In such staggered
mesh, total magnetic flux is identically conserved when we evolve
magnetic fields through Faraday's equation.
%Moreover, the divergence-free
%condition, if initially satisfied, will always be satisfied.

At $t=0$, we populate our 2D grid with a particular force-free
magnetostatic field configuration $B_z(i,j;t=0)$ that can be
anything (i.e., zero, dipole-like of any direction and strength,
quadrupole-like, etc.). Our aim is to impose a particular
horizontal electric field distribution that will asymptotically
evolve $B_z$ toward a target magnetic field configuration
$B_T(i,j)$. We chose the following procedure: given the
photospheric magnetic field distribution $B_z(i,j;t)$ at each time
step $t$, we scan the full photospheric grid $(i,j)$, and at each
cell position we add four electric field components at the four
edges of the cell such that
\begin{eqnarray}
E_x(i,j;t) & \rightarrow & E_x(i,j;t)-f[B_T(i,j)-B_z(i,j;t)]\ ,\nonumber \\
E_y(i,j;t) & \rightarrow & E_y(i,j;t)+f[B_T(i,j)-B_z(i,j;t)]\ ,\nonumber \\
E_x(i,j+1;t) & \rightarrow & E_x(i,j+1;t)+f[B_T(i,j)-B_z(i,j;t)]\ ,\nonumber \\
E_y(i+1,j;t) & \rightarrow & E_y(i+1,j;t)-f[B_T(i,j)-B_z(i,j;t)]\
, \label{scanning}
\end{eqnarray}
as shown in Fig.~\ref{2Dalgorithm}a. Here $f$ is a free positive
numerical factor used to adjust the rate of convergence. We chose
$f=0.5$, as convergence becomes numerically unstable for $f>0.5$
and too slow for $f<0.5$. At each time step $t$, we start with
$E_x(i,j;t)=E_y(i,j;t)=0$ everywhere, and then scan the 2D grid
and update electric fields according to eqs.~(\ref{scanning}). If
we only had one computational grid-cell, the $z$ component of
Faraday's equation (eq.~\ref{M2}) would be written in a
first-order discretized form as follows:
\begin{equation}
\frac{\partial B_z(t)}{\partial t}=\frac{4cf[B_T-B_z(t)]}{\delta}\
.\label{discreteFaraday}
\end{equation}
This equation would be integrated numerically to obtain the
asymptotic solution $B_z(t\rightarrow\infty)=B_T$. When the
algorithm is applied sequentially to the full horizontal grid,
though, electric fields on edges that correspond to neighboring
cells will be updated twice as the algorithm is applied in both
cells. In the sketch of Fig.~\ref{2Dalgorithm}b-f, one can see how
the electric fields of neighboring cells are updated for a simple
magnetic field distribution consisting of three plus three
neighboring cells with equal and opposite magnetic fields. An
important point to make here is that the process can be
generalized in spherical coordinates (a step to take in the
future) besides the Cartesian implementation shown here.

The above 2D procedure can be directly generalized in 3D. We use a
cubic cartesian grid in order to avoid numerical artifacts
associated with the axis of a spherical or cylindrical coordinate
system. Note that in such a grid, the photosphere nowhere
coincides with any of the grid cell phases. We define as our solar
boundary surface the outer surface of the largest cartesian volume
lying entirely within $r_{\odot}$. Given a target photospheric
radial magnetic field distribution
$B_T(r_{\odot},\theta,\phi)\hat{\bf r}$ in heliocentric spherical
coordinates (see the discussion in \S~4 on how $B_T$ is obtained
from a Carrington magnetogram), we define the target magnetic
field at the center of every boundary cell surface at positions
$(x,y,z)$ as the projection perpendicular to that surface of the
field $B_T(r_{\odot},\theta,\phi) (r_{\odot}/r)^2\ \hat{\bf
r}\equiv (B_{Tx},B_{Ty},B_{Tz})$ (for practical purposes, it is
assumed here that the magnetic field inside a certain depth below
the photosphere is purely radial, since our cartesian grid does
not trace too accurately the solar surface). As before, we then
scan the boundary surface and update electric fields as
\begin{eqnarray}
E_x(i,j,k;t) & \rightarrow & E_x(i,j,k;t)-f(B_{Tz}(i,j,k)-B_z(i,j,k;t))\ ,\nonumber \\
E_y(i,j,k;t) & \rightarrow & E_y(i,j,k;t)+f(B_{Tz}(i,j,k)-B_z(i,j,k;t))\ ,\nonumber \\
E_x(i,j+1,k;t) & \rightarrow & E_x(i,j+1,k;t)+f(B_{Tz}(i,j,k)-B_z(i,j,k;t))\ ,\nonumber \\
E_y(i+1,j,k;t) & \rightarrow &
E_y(i+1,j,k;t)-f(B_{Tz}(i,j,k)-B_z(i,j,k;t))\ ,\nonumber \\
\nonumber \\ \nonumber
E_y(i,j,k;t) & \rightarrow & E_y(i,j,k;t)-f(B_{Tx}(i,j,k)-B_x(i,j,k;t))\ ,\nonumber \\
E_z(i,j,k;t) & \rightarrow & E_z(i,j,k;t)+f(B_{Tx}(i,j,k)-B_x(i,j,k;t))\ ,\nonumber \\
E_y(i,j,k+1;t) & \rightarrow & E_y(i,j,k+1;t)+f(B_{Tx}(i,j,k)-B_x(i,j,k;t))\ ,\nonumber \\
E_z(i,j+1,k;t) & \rightarrow &
E_z(i,j+1,k;t)-f(B_{Tx}(i,j,k)-B_x(i,j,k;t))\ ,\nonumber \\
\nonumber \\ \nonumber
E_x(i,j,k;t) & \rightarrow & E_x(i,j,k;t)-f(B_{Ty}(i,j,k)-B_y(i,j,k;t))\ ,\nonumber \\
E_z(i,j,k;t) & \rightarrow & E_z(i,j,k;t)+f(B_{Ty}(i,j,k)-B_y(i,j,k;t))\ ,\nonumber \\
E_x(i,j,k+1;t) & \rightarrow & E_x(i,j,k+1;t)+f(B_{Ty}(i,j,k)-B_y(i,j,k;t))\ ,\nonumber \\
E_z(i+1,j,k;t) & \rightarrow &
E_z(i+1,j,k;t)-f(B_{Ty}(i,j,k)-B_y(i,j,k;t))\ ,\nonumber \\
\label{scaning3D}
\end{eqnarray}
for the corresponding boundary cells that face in the $z$, $x$,
and $y$ direction respectively. As in 2D, each boundary electric
field component is updated twice from its two neighboring boundary
surface grid cells.

\begin{figure}[t]
\centerline{\includegraphics[width=1.\textwidth,clip=]{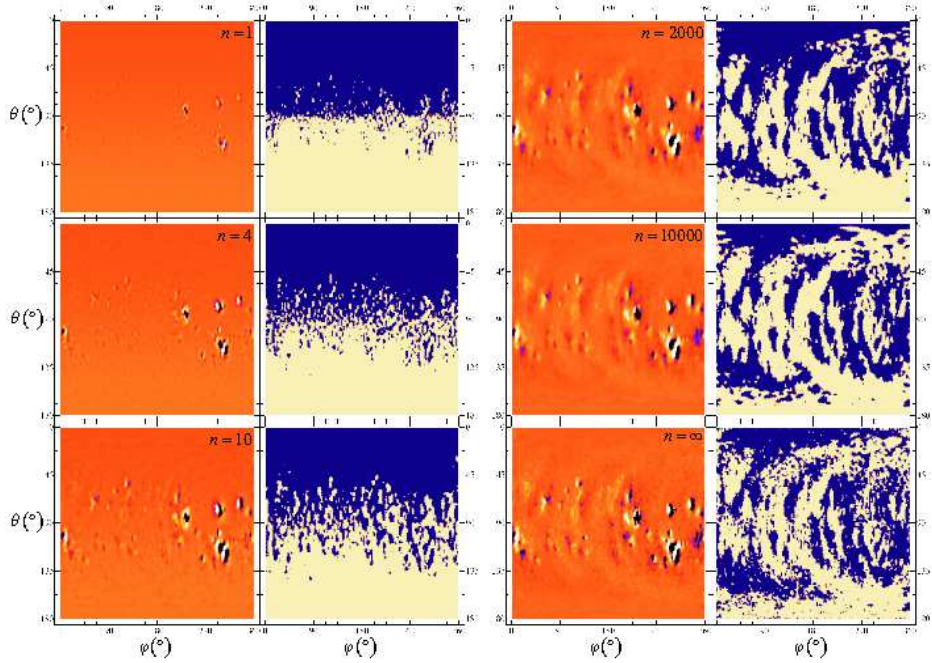}}
\caption{Evolution of the simulated photospheric normal field
component in heliocentric $(\theta,\phi)$ coordinates. Sub-plots
labelled according to time-integration step. The same photospheric
instances are shown with two different color scales. Left: linear
color scale from -1kG to +1kG. Right: brighter areas: $B_r >0$;
darker areas: $B_r <0$. Initial configuration: dipolar, with
$B_r(\theta=0^\circ)=-0.85$~G. Target magnetogram labelled
"$\infty$". This time-sequence results in the equilibrium solution
shown in Fig.~\ref{solar1}.} \label{timesequence}
\end{figure}

To test our algorithm we have used a synoptic magnetogram mainly
referring to Carrington rotation~2009 recorded by the
Michelson-Doppler Imager (MDI; \opencite{scherrer_etal95}) onboard
the Solar and Heliospheric Observatory (SoHO). The choice was
deliberate: Carrington rotation~2009 corresponds to the period of
October-November 2003 (the so-called `Halloween 2003' period) when
the Sun exhibited unusually high eruptive activity and the solar
active-region belt was populated by several very complex active
regions (perhaps an active-region nest, as proposed by
\opencite{zhou_etal07}). Using a solar dipole aligned with the
dipole component of the given magnetogram as the initial
condition, the photospheric magnetic field evolution toward the
target is depicted in Fig.~\ref{timesequence}. Here, the entire
solar disk is represented in $(\theta,\phi)$ heliocentric
coordinates. Photospheric snapshots are labelled by the number $n$
of numerical time-integration steps, with the target magnetogram
denoted with an infinite iteration number ($n=\infty$). One may
notice that the field evolution mimics flux emergence/submergence
through the solar photosphere. Our procedure quickly reproduces
the complex active regions present in the disk, but takes much
longer to reproduce the weaker, large-scale field distribution in
the polar regions.

At this point, we should discuss the spatial resolution of our
computational grid. Ideally we would like our uniform cubic grid
to be able to capture both the details of the `small-scale',
strong active-region magnetic field and the large-scale structure
of the much weaker coronal field. The first requirement calls for
a size element $\delta$ comparable to the resolution of the
SoHO/MDI synoptic magnetogram, that is, $\delta \sim 0.01
r_{\odot}$ or $0.5^\circ$ per pixel. Coupled with the second
requirement that dictates a spatial extent of $\sim 1.5 r_{\odot}$
from the photosphere, it would give rise to a computational domain
of $\sim 10^8$ cubic elements. For our calculation, that is
performed on a typical desktop workstation, this grid size is
infeasible. We thus reduced the original resolution of the
synoptic magnetogram by a factor of 4 ($\delta \sim 0.04
r_{\odot}$ or $\sim 2^\circ$ per pixel). If we were extrapolating
for only a part of the solar disk, say, an active region, this
restriction would be much less stringent, of course. As a result,
the target synoptic magnetogram in Fig.~\ref{timesequence} is
shown with resolution degraded by a factor of 4. Although we
cannot be certain about the degradation factors of other methods,
say, the magnetostatic or magnetohydrostatic methods of T.
Wiegelmann and colleagues, significant spatial degradation is,
indeed, a necessity that one has to deal with when calculating the
global solar magnetic field. The numerical time-integration step
of our simulation is $0.5\delta/c=0.05$~sec. Our simulations need
about 24~hours on a desktop Intel i7 quad-core workstation to
complete 10,000 time-integration steps.
%Given our current spatial grid resolution,
%this represents an electrodynamic time-evolution interval of about
%8 min.

\begin{figure}
\centerline{\includegraphics[width=1.\textwidth,clip=]{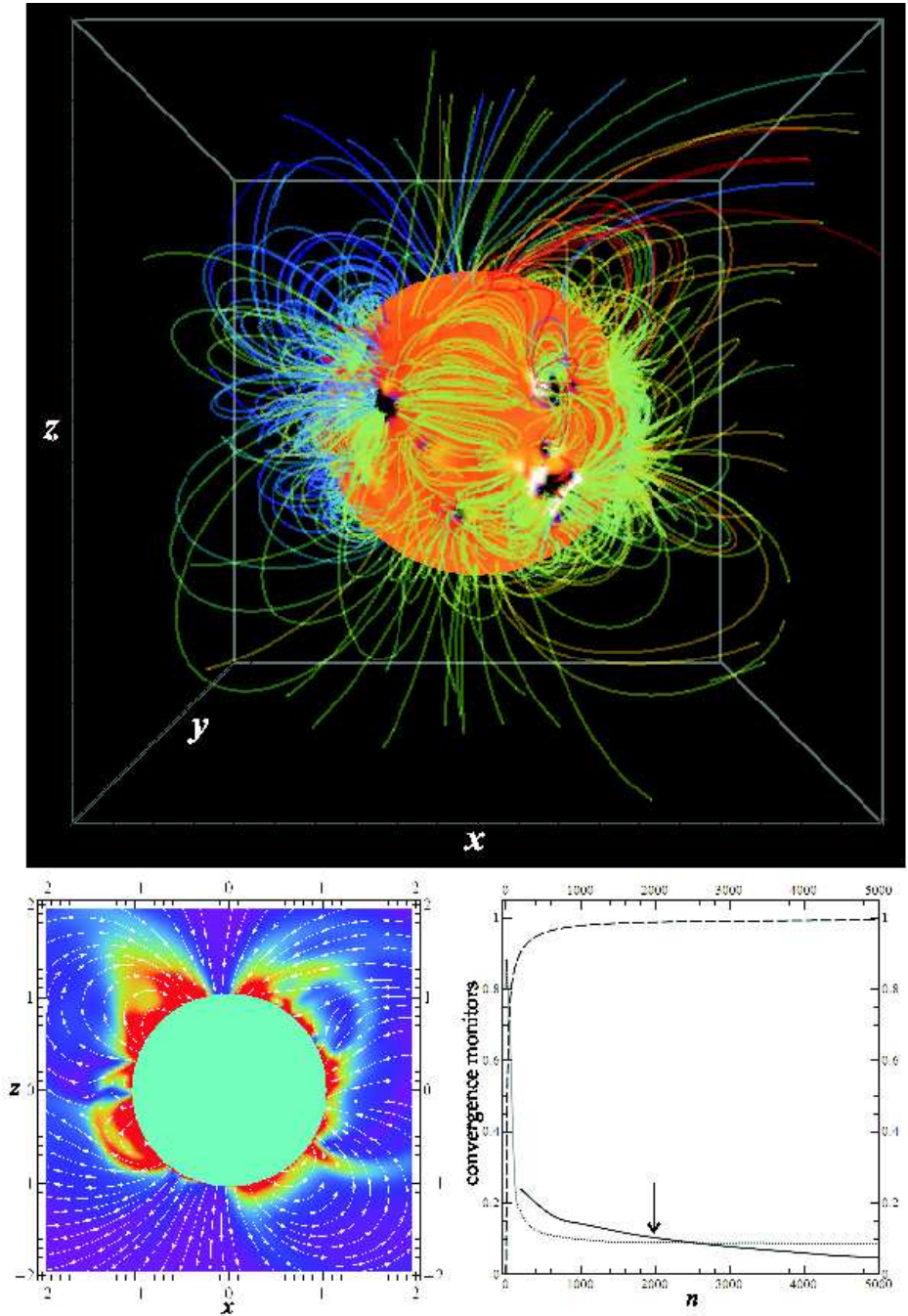}}
\caption{Global nonlinear force-free coronal configuration
corresponding to the target photospheric radial field of
Carrington rotation 2009. Initial dipole polar magnetic field:
$B_r(\theta=0^\circ)=-0.85$~G. Top panel: 3D configuration of
magnetic field lines as viewed from the direction $\phi=0^\circ$.
Field-line color: twist parameter $\alpha\equiv (\nabla\times {\bf
B})/{\bf B}$. Photospheric color scale as in
Fig.~\ref{timesequence}. Lower left panel: final configuration of
magnetic fields projected onto a 2D meridional cut along
$\phi=90^\circ$. Color scale: absolute value of the electric
current $J$. Lower right panel: evolution of convergence monitors
with numerical time-integration step. Dashed line: magnetic field
energy normalized to its asymptotic value attained after 10,000
steps. Dotted line: average weighted sine of the angle between
${\bf J}$ and ${\bf B}$. Solid line: average absolute difference
between target and actual magnetograms over average absolute
target magnetic field. Arrow: numerical integration step that
corresponds to the plotted solution (this is when all convergence
monitors fall below 10\% of their asymptotic value). }
\label{solar1}
\end{figure}
\begin{figure}
\centerline{\includegraphics[width=1.\textwidth,clip=]{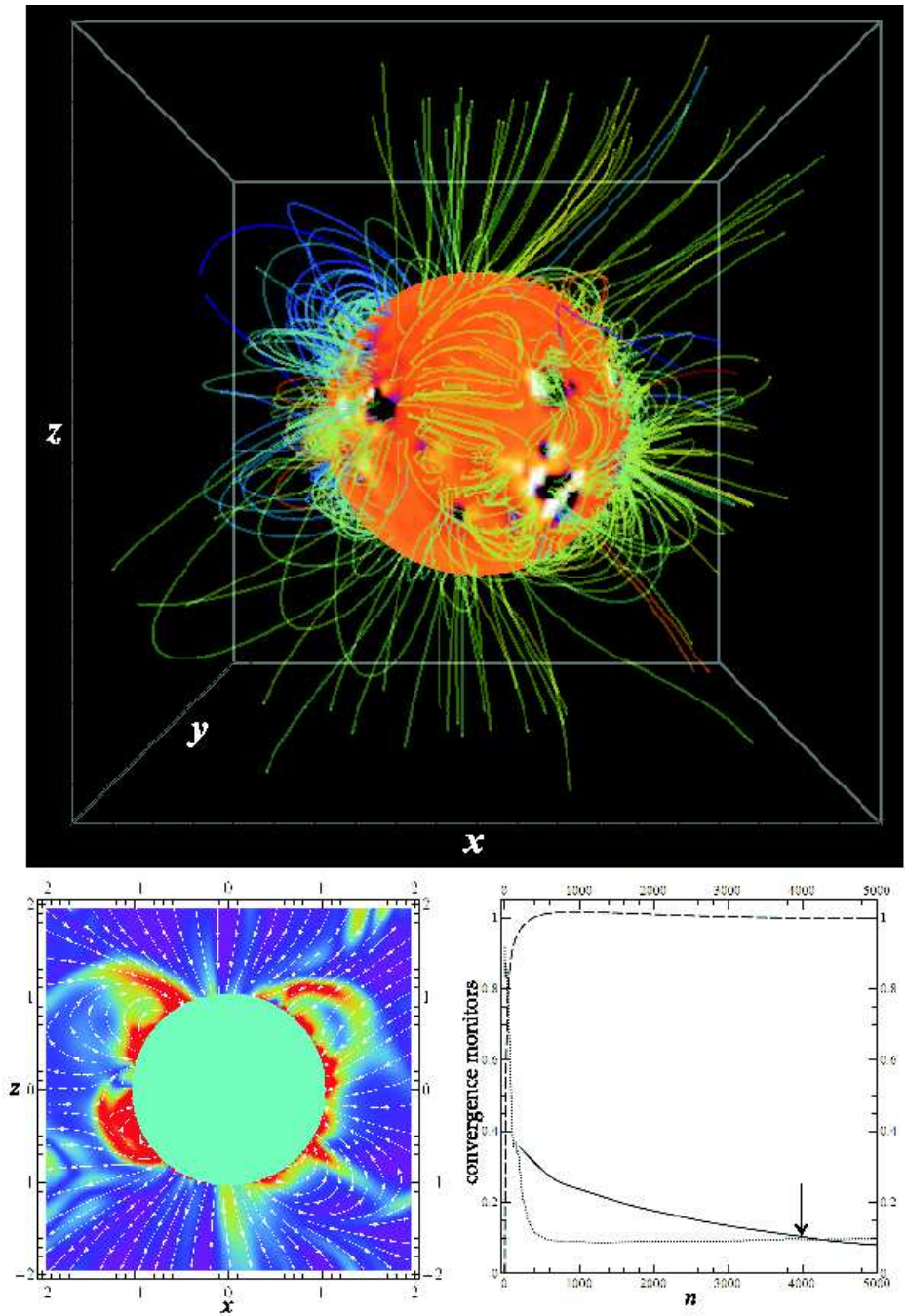}}
\caption{Same as figure~\ref{solar1} but with initial dipole polar
magnetic field $B_r(\theta=0^\circ)=-2$~G. 3D field lines shown
have the same footpoints as in Fig.~\ref{solar1}.} \label{solar2}
\end{figure}
\begin{figure}
\centerline{\includegraphics[width=1.\textwidth,clip=]{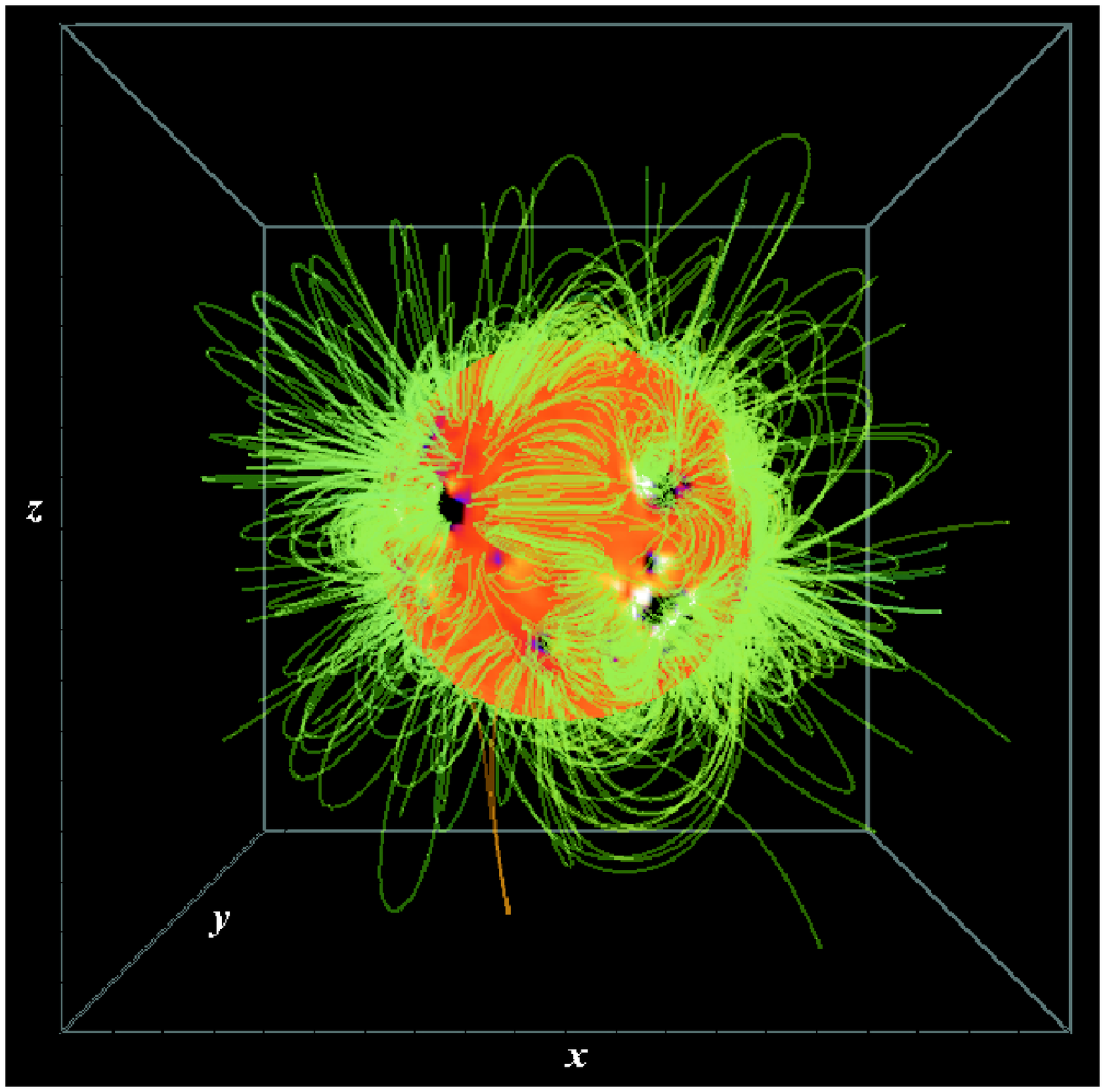}}
\caption{Coronal configuration with nearly zero electric currents
(potential) that corresponds to the same target photosphere as in
figs.~\ref{solar1} \& \ref{solar2}, obtained after 10,000 time
integration steps.} \label{potential}
\end{figure}
\section{The coronal magnetic field}
\label{S-corona}
Given the above boundary condition for the photospheric electric
field we now follow the respective magnetic field evolution in the
corona. This is treated as a pure electrodynamic problem that can
be described through the formalism of FFE. We emphasize, though,
that this is not a vacuum electrodynamic problem since electric
currents and electric charges are allowed to develop everywhere in
the corona to comply with the ideal MHD condition,
eq.~(\ref{idealMHD}).

Electrodynamic waves will traverse the computational region
informing the corona that the photosphere is changing. As they do
so, they leave behind a modified coronal electric and magnetic
field distribution. Similar waves are continuously generated
during photospheric magnetic flux emergence/submergence and travel
out to larger heights in the corona to be absorbed in the non
force-free region of the solar wind. This absorption is
implemented numerically through a Perfectly Matched Layer
(hereafter PML; \opencite{berenger94}, \citeyear{berenger96}; see
also

\noindent \opencite{kc09}), that is,  a non-reflecting, perfectly
absorbing boundary that mimics open space.

As the distribution of the radial photospheric field approaches
that implied by the target magnetogram, the photospheric electric
fields will gradually fade. As a consequence, the coronal
electrodynamic waves and electric fields will also fade, and the
coronal magnetic field will gradually approach a force-free
magnetostatic configuration satisfying the ideal MHD condition. In
doing so, a nonlinear coronal current network develops. In
Figs.~\ref{solar1}-\ref{solar2}, we show two examples of global
coronal solutions reached for two different initial coronal field
configurations. In the lower right panels we plot the evolution
with time-integration step of i) the total magnetic energy $\int
B^2 {\rm d}V$ over our computational volume $V$ normalized to its
asymptotic value obtained after $n=10,000$ time-integration steps
(dashed line), ii) the average sine of the angle between the
electric current ${\bf J}\propto \nabla\times {\bf B}$ and the
magnetic field ${\bf B}$ weighted by the magnitudes of ${\bf J}$
and ${\bf B}$ (dotted line), and iii) the average absolute
difference between the target and actual magnetograms normalized
to the average magnitude of the target photospheric magnetic field
(solid line). As can be seen, the first two monitors approach
numerical convergence to within 10\% in about 500 time-integration
steps, but the third monitor takes roughly ten times as many steps
to drop below the level of 10\%. Indeed, the photospheric magnetic
field needs longer integration times to settle to its asymptotic
value, implying that surface electric fields take longer times to
fade out. Currently we consider as equilibrium solution the state
in which all convergence monitors fall within 10\% of their
asymptotic value. This practice will be reconsidered in future
efforts. In the lower left panels we show the final configuration
of magnetic fields projected onto a 2D meridional cut along
$\phi=90^\circ$, colored by the magnitude of the field-aligned
electric current. In the top panels we show a 3D configuration of
magnetic field lines as observed from the direction
$\phi=0^\circ$, colored according to the twist parameter
$\alpha\equiv (\nabla\times{\bf B})/{\bf B}$ (magnetic field line
color is redish where ${\bf J}$ is parallel to ${\bf B}$, blueish
where ${\bf J}$ is anti-parallel to ${\bf B}$, and light green
where ${\bf J}\approx 0$).

In Fig.~\ref{solar1}, we begin with a dipole magnetic field
counter-aligned with the solar axis with
$B_r(\theta=0^\circ)=-0.85$~G, and aim for an actual global
photospheric magnetic field obtained through part of Carrington
magnetogram~2009. The chosen initial polar magnetic field value
corresponds to the average one in the given magnetogram. By
choosing this particular value we `assist' the longer-term
numerical convergence of our code. As can be seen in
Fig.~\ref{solar1}, the target magnetogram is reproduced to within
10\% in about 2,000 time integration steps (see also time sequence
of Fig.~\ref{timesequence}). This is when we plot the global
coronal solution shown in Fig.~\ref{solar1}. The mean weighted
sine of the angle between the electric current and the magnetic
field vectors is $\leq 0.1$, and the values of $|\nabla\cdot {\bf
B}|/[(\frac{\partial B_x}{\partial x})^2+(\frac{\partial
B_y}{\partial y})^2+(\frac{\partial B_z}{\partial
z})^2]^{\frac{1}{2}}$ are on the order of $10^{-5}$. The $\alpha$
parameter values are in the range $\sim \pm 10^{-3}\
\mbox{Mm}^{-1}$ (3 standard deviations away from its zero mean
value). Notice that such twist is not unreasonable in the solar
corona. Here we should clarify that, although the photosphere has
relaxed to its equilibrium configuration, continuing numerical
integration beyond this stage results in a gradual disappearance
of coronal electric currents due to the numerical diffusion
inherent in any numerical code such as this one.  In other words,
coronal currents are continuously `eroded' by numerical
diffusivity. We know that, by the time we manage to reproduce our
target photospheric boundary condition to within 10\%, some amount
of the coronal electric current will be lost due to numerical
diffusivity. Beyond that point, the corona evolves through a
sequence of magnetostatic equilibria, each with smaller amount of
electric current. After about 10,000 time integration steps, we
reach a configuration (Fig.~\ref{potential}) with very small
coronal electric currents ($\alpha\sim \pm 10^{-4}\
\mbox{Mm}^{-1}$). This current-free configuration is uniquely
determined for a closed volume or for a lower boundary condition
and the semi-infinite space above it (e.g., \opencite{aly_87}),
which is our approximation.

Figure~\ref{solar2} is similar to Fig.~\ref{solar1}, only the
initial magnetic field is dipolar counter-aligned with the solar
axis with $B_r(\theta=0^\circ)=-2$~G. As before, we reproduce very
quickly the main photospheric active regions, but now the polar
photospheric regions take longer to form, within about 4,000 time
integration steps. One might expect that, due to the slower
convergence, the resulting coronal electric currents are now
slightly weaker than before ($\alpha$ values of the order $\pm
0.75\times 10^{-3} \mbox{Mm}^{-1}$), and the two coronal
configurations are somewhat different. If we continue the time
integration for much longer times, coronal electric currents will
also fade gradually, and the coronal configuration will approach
the same potential configuration shown in Fig.~\ref{potential}.
%\begin{figure*}
%\begin{center}
%\includegraphics[width=15cm]{solar2.eps}
%\end{center}
%\caption{Global nonlinear force-free coronal configuration with
%dipole photospheric magnetic field. Zero initial field. a)
%evolution of various convergence monitors (solid line: re-scaled
%magnetic field energy; dotted line: average weighted sine of the
%angle between ${\bf J}$ and ${\bf B}$; dash-dotted: difference
%between actual and target photospheric magnetograms) during the
%approach to the final solution. b) final configuration of magnetic
%fields projected onto a 2D meridional cut along $\phi=90^\circ$. Color
%scale: the magnitude of ${\bf J}$. c) 3D configuration of magnetic
%field lines as viewed from the direction $\phi=0^\circ$, Color scale:
%the dimensionless field-aligned parameter $\alpha\equiv
%(\nabla\times {\bf B})/{\bf B}$ in units of inverse Mm.}
%\label{dipole1}
%\end{figure*}

%\begin{figure*}
%\begin{center}
%\includegraphics[width=15cm]{solar2.eps}
%\end{center}
%\caption{Same as figure~\ref{dipole1} but with a dipole initial
%global magnetic field configuration 2 times as strong as and
%aligned with the one that corresponds to the target dipole
%photospheric configuration.} \label{dipole2}
%\end{figure*}

We also tested our algorithm in a much simpler configuration,
namely that of a solar dipole without magnetic flux in the
active-region belt. We used various initial configurations (zero
or dipolar). In all cases, the equilibrium solution is almost
indistinguishable from the vacuum magnetostatic dipole (potential
field), with very small field-aligned electric currents ($\alpha$
values of the order $\pm 10^{-4} \mbox{Mm}^{-1}$). It seems that
the absence of magnetic flux in the active-region belt gives rise
to weak coronal electric currents and by the time the photosphere
relaxes to its given boundary configuration numerical diffusivity
has enough time to dissipate them effectively. In all cases, the
average weighted value of the sine of the angle between the
electric current and the magnetic field is $< 0.1$, and the
maximum value of $|\nabla\cdot {\bf B}|/[(\frac{\partial
B_x}{\partial x})^2+(\frac{\partial B_y}{\partial
y})^2+(\frac{\partial B_z}{\partial z})^2]^{\frac{1}{2}}$ is of
the order $10^{-5}$. We obtained similar results when we
considered different dipole orientations with respect to the axes
of our numerical grid. This is evidence that our results are grid
independent.

\section{Discussion and conclusions}
\label{S-final}
Both solutions shown in Figs.~\ref{solar1} and \ref{solar2}
represent force-free magnetostatic configurations satisfying the
ideal-MHD condition that correspond to the same magnetogram on the
solar photosphere, therefore, they both represent solutions to our
problem. The potential solution shown in Fig.~\ref{potential} also
represents a solution to our problem. Nature, however, chooses
only one particular configuration. Our numerical approach suggests
that our reconstructed field retains some memory of the prehistory
of its evolution: it depends on the initial global configuration
and on the course toward equilibrium. This is analogous to what we
suspect happens in the real Sun as well, when different solar
cycles succeed one another. If the target boundary does not have
any well-formed active regions, i.e. like in a solar-minimum
configuration, our simulation results in the unique
potential-field solution between the photosphere and the source
surface or for the semi-infinite space above the photosphere. This
is also what full-fledged magnetohydrodynamic global models give,
in this case approaching the results of potential-field models
\cite{riley_etal06}. Obviously there is no straightforward answer
to the question of how to choose one particular coronal field
configuration from another. One way could be to introduce
additional control monitors such as magnetic free energy in our
computational volume or the Poynting flux through the photospheric
boundary. Another way would be to make use of full vector
magnetograms whenever those are available (for example, in
eqs.~(\ref{scaning3D}) we could involve the three magnetic field
components on the boundary toward the target components $(B_{Tx},
B_{Ty}, B_{Tz})$ of the observed field vector ${\bf B}$; not only
its radial component $B_r\hat{\bf r}$). This major issue, that
could fully constrain our solution, will be addressed in a
forthcoming publication.

Finally, we address a problem associated with the observational
method via which the radial component of the photospheric magnetic
field $B_T(r_{\odot},\theta,\phi)$ is inferred from measurements
of the line-of-sight magnetic field
$B_{LOS}(r_{\odot},\theta,\phi)$. We will only consider here the
method applied to a Carrington synoptic magnetogram, although our
discussion below also applies to local observations of particular
photospheric active regions, as well. In a Carrington magnetogram
we are observing the photosphere along solar meridional sections
that contain our line of sight. As the Sun rotates, the full
photosphere is covered and a global magnetogram is obtained under
the assumption that the magnetic field structure remains unchanged
during a full solar revolution (obviously, this is not quite true,
but is the best one can do with present day satellite technology,
avoiding also further interference with observational data). In
order to obtain the radial magnetic field, a further assumption is
made, namely that the magnetic field at the base of the corona is
everywhere perpendicular to the photosphere, i.e.,
\begin{equation}
B_T(r_{\odot},\theta,\phi)=
\frac{B_{LOS}(r_{\odot},\theta,\phi)}{\sin\theta}\ .
\label{deprojection1}
\end{equation}
This second assumption is also not totally accurate, but is the
best one can do without knowledge of the photospheric field
vector. The orientation of this field can be inferred from a full
3D reconstruction of the global coronal magnetic field, and this
is precisely what we are doing in the present work. We, therefore,
propose to include in a future publication one more step in the
iteration procedure implemented along the photosphere, namely,
\begin{enumerate}
\item At time $t=0$, start with a purely radial target for the
normal photospheric field as given by eq.~(\ref{deprojection1}).
\item Apply the numerical integration steps and calculate the
interim line-of-sight component $B_{LOS(num)} (r_{\odot}, \theta,
\phi; t)$. In general, this will be different from

\noindent $B_r (r_{\odot}, \theta, \phi; t) sin\theta$
%(see eq.~\ref{deprojection1})
because a nonzero horizontal magnetic
field now develops in the photosphere. \item Rescale the target
field $B_T (r_{\odot}, \theta, \phi)$ by a factor reflecting the
development of horizontal fields on the boundary, i.e.,
\begin{eqnarray}
B_T(r_{\odot},\theta,\phi) & \rightarrow &
\frac{B_{LOS}(r_{\odot},\theta,\phi)}{\sin\theta}\cdot
\left(\frac{B_r(r_{\odot},\theta,\phi;t)\sin\theta}
{B_{LOS(num)}(r_{\odot},\theta,\phi;t)}\right)\nonumber\\
& = & B_{LOS}(r_{\odot},\theta,\phi)\cdot\left(
\frac{B_r(r_{\odot},\theta,\phi;t)}{B_{LOS(num)}(r_{\odot},\theta,\phi;t)}\right)
\ . \label{scaling}
\end{eqnarray}
Here, the geometric term $sin\theta$, reflecting the assumption of
purely radial field on the boundary in eq.~(\ref{deprojection1}),
is replaced by the more general term $B_{LOS(num)} / B_r$.
\end{enumerate}

The above iterative procedure, implemented together with
eqs.~(\ref{scaning3D}), continuously redefines the target radial
photospheric field, $B_T$, gradually relaxing the assumption of
purely radial fields in the photosphere. To our knowledge, this is
the first time that the full 3D coronal field geometry has been
taken into account in de-projecting the photospheric magnetic
field.

The quest toward choosing the most realistic global reconstruction
of the solar coronal magnetic field at the lowest possible
computational expense is, obviously, still ongoing. Despite the
non-uniqueness of our solution we believe that the proposed method
is interesting and may even turn out to become promising, due to
the following two reasons: firstly, ours is the first technique
that reaches a nonlinear force-free solution using only the radial
photospheric field that is readily available from various sources.
Attempting to match our force-free tangential evolution in the
photosphere with the tangential component provided by SOLIS/VSM or
will be provided by SDO/HMI full-disk vector magnetograms may
further constrain, or determine, hopefully, our solution.
Secondly, our method alleviates the need for preprocessing the
forced photospheric fields to better comply with a coronal
force-free solution. Preprocessing introduces additional
assumptions (e.g., \opencite{tadesse_etal09}) so the fact that we
asymptotically reach the target synoptic magnetogram at the same
time ensuring a force-free solution above each interim boundary
configuration may prove advantageous. Concerning our
non-uniqueness problem, we will consider, classify, examine, and
implement various revisions that may potentially present an
opportunity to improve our global solar magnetic field
reconstruction method.

%
%%%%%%%%%%%%%%%%%%%%%%%%%%%%%%%%%%%%%%%%%%%%%%%%%%%%%%%%%%%%%%%%%%%%%%%%%%%
\begin{acks}
This work uses synoptic solar magnetograms from SoHO/MDI. SoHO is
a project of international cooperation between ESA and NASA.
\end{acks}
%
%%% BIBLIOGRAPHY %%%%%%%%%%%%%%%%%%%%%%%%%%%%%%%%%%%%%%%%%%%%%%%%%%%%%%%%%%%
%
\bibliographystyle{spr-mp-sola}
\bibliography{SoPh_references}
\end{article}
\end{document}